# Dispersed Indeterminacy

Moses Fayngold

*Department of Physics, New Jersey Institute of Technology, Newark, NJ 07102*

A state of a single particle can be represented by a quantum blob in the corresponding phase space, or by a cell in its 2-D subspace. Its area is frequently stated to be no less than one half of the Plank constant, implying that such a cell is an indivisible quantum of the 2-D phase space. But this is generally not true, as is evident, for instance, from representation of some states in the basis of innately discrete observables like angular momentum. Here we consider some *dispersed states* involving the evanescent waves (EW) different from that in the total internal reflection. Such states are represented by a set of separated point-like cells, but with a large total indeterminacy. An idealized model has a discrete Wigner function forming an infinite periodic array of dots on the phase plane. The question about the total momentum indeterminacy in such state is discussed. We argue that the transverse momentum eigenstates corresponding to the considered EW-s cannot be singled out by any known measurement procedure, and the whole infinite set of the corresponding eigenvalues can contribute only a certain fraction to the observed momentum indeterminacy which remains finite.

## 1. Introduction

Due to quantum indeterminacy, a state of a single particle can be represented in its phase space $(p, q)$ by a cell of area

$$\Delta A \geq \frac{1}{2}\hbar \qquad (1.1)$$

(we will use here the term "phase space" to refer to a 2-D subspace representing one degree of freedom $q$). Condition (1.1) reflects the impossibility of a *single* sharp spike in the Wigner function (WF) [1] corresponding to a given state [2-4]. Singular cells with minimal possible size $(1/2)\hbar$ represent the *coherent states* which are considered as being closest to the corresponding classical states [4, 5]. In particular, the cell representing a coherent state, can move along the trajectory prescribed by Classical Mechanics (CM) while retaining its size and shape.

Thus, with respect to (1.1), the phase space can be considered as quantized (granulated). But indeterminacy restricts only the minimal size of cells, not their shape which can be, e.g., circular or elliptical (*squeezed coherent states* [7-9], Fig. 1a ). Its leniency in this respect goes as far as to allow a single cell to be stretched into a line (Fig. 1 b, c).

The WF of a pure state $|\Psi\rangle$ in the $x$-representation $\langle x|\Psi\rangle = \Psi(x)$ is [2-6]

$$W(x, p_x; t) = \frac{2}{h} \int \Psi^*(x-x', t)\Psi(x+x', t) e^{-2ik_x x'} dx', \quad k_x \equiv \frac{p_x}{\hbar} \qquad (1.2)$$

This definition, if applied to a position or momentum eigenstate, leads to the corresponding WF-s which are discrete along $x$- or $p_x$-dimension, respectively. They do not possess some basic properties of "regular" WF. In particular, their modulus may be arbitrarily large, in contrast with the rule $|W(x, p_x)| \leq 2/h$ for regular WF. For example, taking the momentum eigenstate $|p'_x\rangle$ in the $x$-representation (de Broglie's wave)

$$\Psi_{p'_x}(x) = \langle x|p'_x\rangle = \frac{1}{\sqrt{2\pi}} e^{ik'_x x} \qquad (1.3)$$

and putting it into (1.2) gives

$$W_{p'_x}(x, p_x) = \frac{2}{h} \delta(k_x - k'_x) \qquad (1.4)$$

The same result obtains if we use the $p_x$-representation $\mathcal{F}_{p'_x}(p_x) = \langle p_x|p'_x\rangle = \delta(p_x - p'_x)$ of the same state, which is just the Fourier transform of (1.3). The result is $x$-independent, thus representing correctly the basic features of the original state (1.3) – the probability distribution which is uniform in the coordinate space and sharply localized in the momentum space. But at the same time it is dimensionally different from regular WF and has an infinite magnitude. Both distinctions, as indicated below, are due to a specific normalization of de Broglie's states.

In case $q \equiv x$, Fig. 1b shows an infinite baseline of the WF (1.4). Similarly, Fig. 1c shows the baseline of the WF of state $\Psi_{x'}(x) = \delta(x - x')$ which is the eigenfunction of position operator corresponding to eigenvalue $x'$.

It is claimed sometimes that the eigenstates of the position or momentum operator, whose WF are the $\delta$-functions with an infinite baseline in the phase space, cannot represent a physical state [3]. The argument used to prove this claim is that:
1) a line has no area and  2) the mentioned eigenstates are not members of the Hilbert space.

The first argument ignores the fact that a line in question is a limit of a finite cell stretched under condition of conserving its area $\Delta A$. The resulting line is of the same nature as the graph of a $\delta$-function, e.g. $\delta(x - x')$. While such a graph is merely a line with the point base, it has the *nonzero* (unit) area according to a fundamental property of $\delta$-function $\int \delta(x - x')dx = 1$ for any integration region containing $x'$.

The second argument ignores the fact that mutually orthogonal eigenvectors $|\psi_\alpha\rangle$ of an operator with continuous spectrum of eigenvalues $\alpha$ form an innumerable set; and accordingly, they are elements of the Hilbert space with innumerable dimensionality. The distinction of such a space is a "continuous" (depending on *continuous* parameter $\alpha$) normalization condition $\langle \psi_\alpha | \psi_{\alpha'} \rangle = \delta(\alpha - \alpha')$, instead of $\langle \psi_n | \psi_{n'} \rangle = \delta_{nn'}$ for a discrete set [10, 11]. In all other respects it retains all basic properties of "regular" Hilbert spaces.

Thus, the first argument is generally not correct, and the second one implies only a very special subset of the Hilbert spaces; therefore neither argument can prove the quoted statement.

It would be more accurate to say that eigenvectors $|p_\alpha\rangle$ or $|x_\alpha\rangle$ are the idealized models of some real states. Most of known representations of physical reality are idealized models, like a point mass in Newton's gravitation law or inertial reference frame in Relativity [12, 13]. Neither of them exists as a real object, but this does not preclude their routine use in Physics. Similarly, the eigenstates of position or momentum operator, while being only the models of certain real states, are indispensable tools in mathematical structure of QM. In this work, we will use the coordinate and momentum eigenstates and their superpositions as the models of some real situations.

## 2. Dispersed states

In the described model, the cells representing the state in the phase space are point-like. In this respect, the corresponding states may appear to be even closer to CM than the coherent states. A possibility to shrink an "indivisible quantum" of the phase space down to a point-like dot seems to be the resurrection of CM within the framework of QM; but the necessity for the corresponding *set* of $\tilde{N} \gg 1$ dots in all such cases takes QM even father away from CM than typical indeterminacy $\Delta p \Delta x \geq (1/2)\hbar$ with continuous $p, x$. In the simplest case we may assume $\tilde{N} = N^2$, where $N$ is the number of dots along either dimension of the phase space. But such a "discrete" phase space is still consistent with restriction (1.1) because we have now a *set of dots* (Fig. 2) in which neither dot can exist alone without all the rest. Even when separated as in Fig. 2, the dots must be considered as *parts of a single* (*albeit dispersed*) *cell*, and their *net* area must satisfy (1.1). All of them together represent the state of a *single* particle.

In known cases with finite *N* there are requirements for *N* to be odd or a power of an odd prime [14-21]. We consider an extreme case with a state represented by an infinite number of point-like dots forming a periodic lattice over the whole phase space (this is one of the cases when some features of a system are maximally pronounced and simple to describe when the number of system's elements is infinite). To be specific, we assume that *q* and *p* in Fig. 2 stand for position and momentum, respectively, so $q = x$. Both – the position and momentum – of each dot is *defined exactly*, and yet the indeterminacy in this case is maximal possible: we have $\Delta p_x \Delta x \to \infty$ since the particle does not know in which dot it "resides" – it is in an equally-weighted superposition of all possibilities. In this case all the space between the dots, while representing the regions where the particle cannot be found in a position or momentum measurement, nevertheless contributes to indeterminacy, making it infinite. Hence the term – dispersed indeterminacy.

The corresponding quantum state may appear to have a purely academic interest, but in fact it can model some real situations like neutron bound states in an atomic chain [22] or a textbook example with photon diffraction through a grating. The latter is a version of diffraction with wave front splitting [23] – just the extension of the double slit experiment to a large number of slits (Fig. 3). An idealized model features a set of $N \to \infty$ infinitely narrow slits separated by a distance *d*. Let the incident light be dimmed to one photon at a time. If there are no attempts to watch the photon in the process, it takes all available virtual paths and then interferes with itself. With the *x*-direction perpendicular to the slits in the screen plane, we can represent the state function (more accurately, electric or magnetic "footprint") of a photon passing through the grating as

$$\psi(x) = \sum_n \langle x | x_n \rangle = \sum_n \delta(x - nd), \qquad |n| = 0, 1, 2, \ldots \qquad (2.1)$$

Each slit $x_n = nd$ is represented by the respective column of dots, which is labeled by *n* in the phase diagram in Fig. 2 (the normalizing factor in a superposition of the type (2.1) is arbitrary and can be dropped).

Following Optics, we call part(s) of the wave front passing through the slit(s) in an opaque screen the "aperture function" [23]. The aperture function (2.1) is represented graphically in Fig. 4a. Its Fourier-transform is

$$\phi(k_x) = \frac{1}{\sqrt{2\pi}} \int \psi(x) e^{-ik_x x} dx = \frac{1}{\sqrt{2\pi}} \sum_n e^{-ik_x nd} \qquad (2.2)$$

Using the identity [24, 25]

$$\sum_m e^{-im\xi} = 2\pi \sum_n \delta(\xi + 2\pi n), \qquad |m| = 0, 1, 2, \ldots, \qquad (2.3)$$

setting $\xi = k_x d$, and applying the rule $\delta(\alpha x) = \alpha^{-1} \delta(x)$, we obtain

$$\phi(k_x) = \frac{\sqrt{2\pi}}{d} \sum_m \delta(k_x - mk_d), \quad k_d \equiv \frac{2\pi}{d} \qquad (2.4)$$

Thus, all possible $k_x$ form a discrete set $k_x^{(m)} = mk_d$ (integer $m$ will number the elements of this set). This is why each slit in succession (2.1) is represented by a respective column of dots in the phase space (Fig. 2), rather than by a continuous vertical line (the latter would be the case, e.g., for a single slit). And by the same token, each value $k_x^{(m)} = mk_d$ corresponding to a particular $m$ is represented by $m$-th row of dots in Fig. 2.

The series of $\delta$-functions on the right in Eq. (2.3) forms the *Dirac comb function*, a term made self-intuitive by its graphical representation in Fig. 4 (it is also known as *Shah function*, *sampling symbol*, or *replicating symbol* [24, 25]). We see that a comb function in the configuration space (Fig. 4a) converts into a comb function in the momentum space (Fig. 4b) and vice-versa.

It is frequently stated that the Gaussian $U(x) = U_0 e^{-\sigma^2/2x^2}$ with its Fourier transform $F(k_x) = U_0/\sigma \, e^{-k_x^2/2\sigma^2}$ is the only function looking the same in both spaces. But this is true only in the set of continuous functions. Fig. 4 shows a state other than Gaussian with two mutually complementary but similarly looking "faces".

Each slit $x_n = nd$ takes its part in formation of *all* waves with $k_x^{(m)} = mk_d$, and conversely, every such wave is generated by the entire succession of slits. Using (2.3, 4), we can write the diffracted wave immediately behind the grating ($z = 0$) as

$$\psi(x) = \frac{1}{\sqrt{2\pi}} \int \phi(k_x) e^{ik_x x} dk_x = \sum_m e^{i m k_d x} \qquad (2.5)$$

We emphasize again that this is a highly simplified description of a real situation. The actual number of slits is, of course, finite, as well as the width $a$ of each slit. The aperture function used to describe the truncated wave front passing through the screen perforations is a very crude approximation of the actual wave function within the slits. This function depends, among other things, on the slit's width and on boundary conditions representing the optical properties of the screen. For a screen of finite thickness $\delta$ with perfectly reflecting surface, each slit works as a non-dissipative optical waveguide with conductive walls, and the wave passing through it can be represented by the corresponding waveguide mode(s). For a sufficiently narrow slit, its threshold frequency exceeds that of the incident light [26], and the passing wave exponentially attenuates along the $z$-direction. It is one of manifestations of the known fact that a photon cannot be squeezed into a stationary state within a region smaller than its wavelength $\lambda$ [27]. As a result, for the slit width $a \ll \lambda$ and a sufficiently thick screen, the amplitude of passing through the screen will be exponentially small. This attenuation precedes the following stage to be considered in Sec. 5 – the exponential decay of the amplitudes as functions of $z$ for sufficiently high $m$, $mk_d > \omega/c$, in a free semi-space outside the screen.

The described model represents the *discrete* limit of a dispersed state, which is shown in Fig. 2. In the considered limit $a \to 0$, $N \to \infty$, each cell shrinks down to a point, but their number becomes infinite, which produces the infinite indeterminacy associated with the whole set of such cells.

### 3. The Wigner function of a dispersed state

Here we compare the "map" of a dispersed state shown in Fig. 2 with its WF right on the transmission side of the screen ($z = 0$). Applying (1.2) to case (2.1) gives

$$W(x, p_x) = \frac{1}{\pi \hbar} \int \sum_m \delta(x - md - x') \sum_n \delta(x - nd + x') e^{-2ik_x x'} dx' =$$
$$= \frac{1}{\pi \hbar} e^{-2ik_x x} \sum_{m,n} \delta(2x - md - nd) e^{2ik_x md} = \quad (3.1)$$
$$= \frac{1}{\pi \hbar d} e^{-2ik_x x} \sum_m e^{2ik_x md} \sum_n \delta\left(2\frac{x}{d} - m - n\right)$$

The time-dependence is dropped from the equation because we consider the stationary state. Applying the rule (2.3) to the last sum on the right transforms (3.1) to

$$W(x, p_x) = \frac{2}{hd} e^{-2ik_x x} \sum_m e^{2ik_x md} \sum_n e^{2ink_d x} \quad (3.2)$$

Now, applying this rule backwards takes us to

$$\sum_m e^{2ik_x md} = \frac{k_d}{2} \sum_m \delta\left(k_x - m\frac{k_d}{2}\right) \quad \text{and} \quad \sum_n e^{2ink_d x} = \frac{d}{2} \sum_n \delta\left(x - n\frac{d}{2}\right) \quad (3.3)$$

Therefore

$$W(x, p_x) = \frac{1}{2\hbar d} e^{-2ik_x x} \sum_{m,n} \delta\left(k_x - m\frac{k_d}{2}\right) \delta\left(x - n\frac{d}{2}\right) = \frac{1}{2\hbar d} e^{-2ik_x x} \sum_{m,n} \delta(\boldsymbol{\sigma} - \boldsymbol{\sigma}_{mn}), \quad (3.4)$$

where $\boldsymbol{\sigma}$ is a 2-vector $\boldsymbol{\sigma} \equiv (k_x, x)$ in the phase plane, and

$$\boldsymbol{\sigma}_{mn} \equiv \left(m\frac{k_d}{2}, n\frac{d}{2}\right) \quad (3.5)$$

Since (3.4) is non-zero only at $\boldsymbol{\sigma} = \boldsymbol{\sigma}_{mn}$, we may write $e^{-2ik_x x} \Rightarrow e^{-imn\pi}$, and finally get

$$W(x, p_x) = \frac{1}{2\hbar d} \sum_{m,n} e^{-imn\pi} \delta(\boldsymbol{\sigma} - \boldsymbol{\sigma}_{mn}) = \frac{1}{2\hbar d} \sum_{m,n} (-1)^{mn} \delta(\boldsymbol{\sigma} - \boldsymbol{\sigma}_{mn}) \quad (3.6)$$

We have come to the above-mentioned apparent paradox: the rules of QM do not allow WF to be a single $\delta$-spike with a point base on the phase plane, but allow it to be a periodic array of such spikes for some discrete states. The value $W(\sigma_{mn})$ is positive if at least one of the numbers $(m, n)$ is even. When both are odd, $W(\sigma_{mn})$ is negative. At points with odd $m$ the sign of $W(\sigma_{mn})$ alternates between "+" ($n$ even) and "$-$" ($n$ odd), and the same holds for rows of points with odd $n$ (Fig. 5). All alternating terms (the ones with either $m$ or $n$ or both odd) are extra – they are absent on the "map" in Fig. 2. But averaging over a region enclosing a few cells effectively eliminates these terms, and the part that survives (with even $m$, $n$) is positive definite. In this respect the WF of the considered dispersed state displays the same behavior as WF for "regular" states described by continuous, square-integrable wave functions.

### 4. "Semi-discrete" Wigner function

We want to emphasize that the phase space of the considered continuous variables with selected discrete subset $x \to x_n = nd$ becomes fully discrete only in the limit $N \to \infty$. For any finite $N$ (and with $a$ vanishingly small) the WF is discrete only along the $x$-dimension, while the $p_x$-dimension remains continuous. The reason is purely mathematical – the relation (2.3) leading to (2.4) does not hold at finite $N$. In the latter case we have a system with the aperture function

$$\Psi_N(x) = \sum_{n=1}^{N} \delta(x - nd) \qquad (4.1)$$

The corresponding WF according to (3.1) is:

$$W_N(x, p_x) = \frac{2}{h} \int \left( \sum_{m=1}^{N} \delta(x - md - x') \right) \left( \sum_{n=1}^{N} \delta(x - nd + x') \right) e^{-2ik_x x'} dx' =$$
$$= \frac{1}{h} \sum_{m,n}^{N} \delta\left( x - (m+n)\frac{d}{2} \right) e^{-2ik_x(x - md)} \qquad (4.2)$$

The mathematical structure of this expression allows us to replace $x \to (m+n)d/2$ in the exponents. This leads to

$$W_N(x, p_x) = \frac{1}{h} \left\{ \sum_{m=1}^{N} \delta(x - md) + \sum_{m \neq n}^{N} \delta\left( x - (m+n)\frac{d}{2} \right) \cos(m-n)k_x d \right\} \qquad (4.3)$$

It is convenient to shift the origin to the middle of the lattice (or grating):

$$x = \tilde{x} + (N+1)\frac{d}{2}, \qquad (4.4)$$

with $\tilde{x}$ being the distance of a chosen point from the new origin. Putting this into (4.3) gives

$$W_N(x, p_x) =$$
$$= \frac{1}{h}\left\{\sum_{m=1}^{N}\delta\left(x-(N-2m+1)\frac{d}{2}\right)+\sum_{m\neq n}^{N}\delta\left(x+(N-m-n+1)\frac{d}{2}\right)\cos(m-n)k_x d\right\} \quad (4.5)$$

(here we dropped the tilde sign on $x$ keeping in mind that $x$ is now a distance from the center of the grating). For $N=1, 2$ and $3$ we have (Fig. 6)

$$W_1(x, p_x) = \frac{1}{h}\delta(x); \quad W_2(x, p_x) = \frac{1}{h}\left\{\delta\left(x+\frac{d}{2}\right)+\delta\left(x-\frac{d}{2}\right)+2\delta(x)\cos k_x d\right\};$$

$$W_3(x, p_x) =$$
$$= \frac{1}{h}\left\{\delta(x+d)+\delta(x)(1+2\cos 2k_x d)+\delta(x-d)+2\left[\delta\left(x+\frac{d}{2}\right)+\delta\left(x-\frac{d}{2}\right)\right]\cos k_x d\right\}$$

## 5. The output state behind the screen

The above-described dispersed states can be observed within the thickness $\delta$ of the totally reflecting perforated screen, when $\psi(x)$ is non-zero only inside the slits. In the limit $a \to 0$ and $\delta \to 0$, the dispersed state will exist only on the plane $z=0$. This does not undermine its importance, since it gives rise to the observable "output state" *behind the screen* ($z>0$), which becomes only more intense at $\delta \to 0$. The output state is much more complicated than the input (incident wave). Albeit not discrete at $z>0$, it is generated by the discrete intermediate state within the screen, and its properties may be crucial for evaluating the indeterminacy $\Delta p_x \Delta x$, therefore we will give here its brief description. To this end, we will use $k_x$, $k_z$ to indicate (in units $\hbar$) the corresponding components of particle's (photon's) momentum.

Assuming that the input state is a monochromatic plane wave with frequency $\omega$ and momentum $\mathbf{k} \parallel \hat{\mathbf{z}}$ (the light incident from below on a horizontal grating), the output state in the semi-space $z>0$ above the grating will be

$$\Psi(\mathbf{r}, t) = \sum_m e^{i\left(k_x^{(m)}x+k_z^{(m)}z\right)}e^{-i\omega t} = \sum_m e^{ik_z^{(m)}z}e^{ik_x^{(m)}x}e^{-i\omega t} =$$
$$= \left\{e^{ikz}+2\sum_{m=1}^{\infty}e^{ik_z^{(m)}z}\cos mk_d x\right\}e^{-i\omega t} \quad (5.1)$$

Here $k = |\mathbf{k}|$, and

$$k_z^{(m)} = \pm\sqrt{k^2-m^2 k_d^2} \quad (5.2)$$

The "$-$" sign in (5.2) corresponds to the output waves on the reflection side of the grating, that is, in the semi-space $z \leq -\delta$. The output there will, apart from the sign of $k$ and $k_z^{(m)}$, generally differ from (5.1) in the superposition amplitudes. We select the "+" sign to focus on the transmission side.

Dropping in (5.1) the temporal factor and taking the limit $z \to 0$, we recover (2.5) or, which is the same, (2.1). Thus, the form (2.1) is just the boundary condition for solution $\Psi(\mathbf{r},t)$ in semi-space $z \geq 0$.

This solution could also be described as a set of cylindrical waves emerging from the respective slits (Fig. 3); but the Cartesian coordinates used here are better suited for description of the phenomenon in the near field (NF). In particular, they show directly that the state *above the screen* is no longer represented by the Dirac comb function. This is immediately seen from (5.1), where the amplitudes of the horizontally propagating waves are different for different *m* at $z > 0$, so the rule (2.3) cannot be applied, and $\Psi(x, z)|_{z>0}$ is not the comb-function. The particle can be found at any *x*, albeit with different probabilities, and we will have, instead of (2.1), the probability distribution shown in Fig. 7: a continuous diffraction pattern with the side maxima becoming more pronounced (at the cost of the initial $\delta$-spikes) with increase of *z*. At sufficiently large *z* we will obtain the familiar picture of diffraction on a grating if, in addition, we truncate its size, that is, take a system with a finite number *N* of slits.

### 5a. Evanescent waves

The interpretation of the output state behind the screen is far from trivial. At sufficiently high *m* the $k_z$ becomes imaginary, $k_z^{(m)} \to i\chi_z^{(m)}$ with real $\chi_z^{(m)} = \sqrt{m^2 k_d^2 - k^2}$, but according to dispersion equation (5.2), the particle's energy *E* remains constant. This converts the corresponding plane wave into an EW similar to that known in the total internal reflection (TIR) [23, 26, 28-31]:

$$\psi_m(\mathbf{r}) = \phi_0 e^{i\left(mk_d x + k_z^{(m)} z\right)} \Rightarrow \phi_0 e^{-\chi_z^{(m)} z} e^{imk_d x}, \quad \chi_z^{(m)} > 0. \qquad (5.3)$$

The transition (5.3) occurs at

$$\left|k_x^{(m)}\right| > k, \quad \text{or} \quad |m| > m_c \equiv [k/k_d], \qquad (5.4)$$

where square-bracketed $[X]$ stands for the integer part of $X$. Instead of heading upward from the grating, each EW propagates clinging to it, with a wavelength shorter than $\lambda$ ($\lambda_m \equiv 2\pi/|k_x^{(m)}| < 2\pi/k$) and accordingly with the slower phase velocity ($u_m \equiv \omega/mk_d < \omega/k$), in view of condition (5.4). For a photon, $u_m < c$, even though the semi-space $z > 0$ is free of any medium. The word "clinging" emphasizes that the wave's amplitude is maximal at the grating ($z = 0$) and exponentially decreases with *z*.

Thus, in contrast with de Broglie's wave of the same frequency, the considered EW-s, while remaining formally the eigenfunctions of the momentum operator, have their vector eigenvalues with a non-zero, imaginary-valued component perpendicular to their propagation direction[1].

With all that, the EW-s described here are essentially different from the EW in TIR, so it is worthwhile to separate them into two distinct varieties. Let us denote EW appearing in TIR as EW1. The EW appearing in the process of passing through a perforated (more generally, non-homogeneous) screen (NHS) will be called EW2. We can indicate three features that distinguish EW2 from EW1.

1. There is only one EW1 for an incident monochromatic plane wave, and only on the transmission side of the interface between the two mediums. In contrast, the EW2 emerge on both sides of the screen and generally have on either side an infinitely broad spectrum of $k_x$, *all corresponding to the same* $\omega$. Such spectrum is generated by any modulation of the wave front of the incident plane wave, and its range is defined by

$$|k_x| > k, \qquad (5.5)$$

which is just (5.4) generalized to continuous $k_x$. While each individual EW2 is maximal at $z = 0$, the whole superposition (5.1) sums up to zero everywhere on the screen outside the slits. Therefore we expect that the experiments using a test particle (e.g., a small polisterene sphere [32]), which have been performed with EW1, will be much more complicated with EW2; in particular, their results must generally depend not only on the particle's coordinate $z$ above the screen, but also on its position *x along the screen*, and the $z$-dependence will not reduce to a single exponent.

2. While a pure unperturbed EW1 is the only actor in the play in the second medium, the EW2 are always accompanied by at least one regular running wave (RW) receding from the screen as seen from (5.1). Generally there forms a system of receding pairs of crossed RW with real $k_z > 0$ (Fig. 8), with the net amplitudes modulated along the *x*-direction as described by the corresponding terms in the last Eq. (5.1); a specific feature of each such pair is a superluminal phase velocity of the resulting "wave front" [33]. To emphasize the fact that EW2 in the NF always come together with at least one RW extending into the far field (FF), we can write generally

$$\Psi(\mathbf{r},\, t) = \left\{ \left( \underbrace{\Psi(\mathbf{r})}_{\text{RW}} \right) + \left( \underbrace{\Phi(\mathbf{r})}_{\text{EW2}} \right) \right\} e^{-i\omega t} \qquad (5.1a)$$

In the special case of the grating considered here this reduces to (5.1) in the form

$$\Psi(\mathbf{r},\, t) = \left\{ \left( \underbrace{e^{ikz} + 2\sum_{m=1}^{m_c} e^{ik_z^{(m)}z} \cos mk_d x}_{\text{RW}} \right) + \left( \underbrace{2\sum_{m>m_c} e^{-\chi_z^{(m)}z} \cos mk_d x}_{\text{EW2}} \right) \right\} e^{-i\omega t} \:, (5.1b)$$

---

[1]The term EW is also applied to a part of a wave function within a potential barrier in 1-D quantum tunneling, but in this case the propagation vector within the barrier has no real components. We also leave out the EW accompanying the guided waves and some surface waves, e.g., on a metal-dielectric interface.

and the similar expression obtains for the reflection side. The existence of the RW will additionally complicate probing EW with other particles, by masking the sought-for effect due to RW scattering from the probe.

At this point, it is convenient to consider the whole process as a scattering problem with the grating as a macroscopic scatterer, similar to the approach used in [34]. In this approach, we describe the output stage as the emergence of the broad *angular* spectrum of **k** with fixed magnitude but ranging from vertical ($\theta = 0$) to horizontal ($\theta = \pm \pi/2$). All output waves separate into 2 subsets – RW ($0 \leq m \leq m_c$, $\theta_m = Arc\sin(mk_d/k)$ - regular diffraction modes), and EW2 ($m > m_c$, $\theta = \pm \pi/2$) (Fig. 8). In a non-monochromatic (non-stationary) state we will have a superposition of expressions (5.1a) with various $\omega$.

3. The third difference is more subtle and yet more fundamental. Any EW1 carries momentum along the interface between the two mediums, which is manifest, e.g., in the Goos-Hänchen effect [26, 28]. Physically, this is a direct consequence of the oblique incidence above the critical angle which is necessary for the emergence of EW1, so *there is no* EW1 *at the normal incidence*. In contrast, the EW2 considered here appear at the normal incidence as in Fig. 3, and in this case they do not carry any *net* momentum along the screen. Accordingly, the output state (5.1) considered as a function of *x* is the system of standing waves. In a more realistic system with finite *N* we will still have the zero net momentum, but with the opposite fluxes on the left and right, carrying energy-momentum along the screen but away from its center as shown in Fig. 8.

Now we will consider some implications of these distinctive properties, especially of No. 3. Let us shift the focus from the *net* momentum to the *individual eigenstates* of $\hat{\mathbf{p}}$ in EW2. An eigenstate with arbitrarily high *m* must carry arbitrarily high momentum $p_x^{(m)} = \hbar k_x^{(m)}$ along the screen. This raises the question: can we observe *separately* each term of superposition (5.1)? In other words, can a momentum measurement preserving the original experimental setup collapse the superposition (5.1) to a single element of the second subset?

Any particle right before its momentum measurement must be effectively free. This condition is satisfied in case of EW1, since the medium in each semi-space and the interface between them are all homogeneous (and precisely because of this we have only one EW1 for each incident plane wave!). Therefore the answer for EW1 is trivial "Yes" – we have the single eigenstate to begin with. The possible experiments with absorption of energy and momentum quanta of EW1 by elementary or compound test particles in the NF were described in [28] (an elementary particle can absorb a EW-quantum because the latter has some tachyonic properties [28]).

The answer for EW2 is much more complicated. Partially, it is prompted by comparison of state (5.1) with a superposition $\Psi_n = \sum_{l,m} c_{l,m} \Psi_{nlm}$ of electron degenerate states $\Psi_{nlm}$ in the Coulomb field. In the latter superposition, each $\Psi_{nlm}$ is a special solution of the stationary Schrodinger equation with the same eigenvalue $E_n$ and *the same boundary conditions*, and can be observed separately. In other words, the energy *E* and the relevant parts of angular momentum **L** are compatible observables in a spherically-symmetrical field. In contrast, no term figuring in (5.1) can be singled out as an individual degenerate state belonging to the same eigenvalue of the Hamiltonian $\hat{H}$. The reason is that none of them *alone* is the special solution satisfying the boundary condition (2.1) – we *must* take the whole set (2.1). Observables $E$ and $p_x$ are incompatible in an *x*-dependent potential describing a NHS. The same can be expressed by noticing that $\hat{H}$ and $\hat{p}_x$-operators have common set of eigenfunctions in EW1 case, but different sets of them in the EW2 case, so that no single $|p_x\rangle$ is an eigenfunction of the *x*-dependent $\hat{H} \to \hat{H}(x)$. This confirms the initial statement that no dot in Fig. 1 can exist separately from the others.

### 5b. Far-field measurements

The measurements in the considered case can be of two different kinds – the FF or NF measurements. In the FF measurements, the above argument is less restrictive, because the field there is less sensitive to the properties of the boundary. This is especially important for the RW-s ($|m| \leq m_c$), which naturally extend into the FF without changes. And in addition, for a grating of any finite size they eventually become spatially separated from each other as shown in Fig. 8. This allows one to observe $\mathbf{p}^{(m)}$ separately for each $|m| \leq m_c$. As seen from Fig. 8, a measurement using, e.g., suitably positioned distant detectors (or a distant observation screen as in diffraction experiments) can collapse the output state to one of the elements of the RW subset, thus completing the measurement. But the elements of the second subset (EW2), as long as they remain attached to the grating thus staying in the NF, do not spatially separate from each other.

The FF measurement of EW2 could be performed on a particle that has slid from the grating, and a measuring device must be sufficiently far from grating's edge, as shown in Fig. 8. In that region, evanescence disappears – asymptotically, all waves will be solutions of a wave equation for a free particle. Accordingly, they all will convert and merge into a single wave with the same wavelength $\lambda$ and phase velocity $u$ as in the input state (some features of such conversion to regular wave are described in [35]). For all $|m| > m_c$, such wave may be close to cylindrical (but not axially-symmetric) wave with the effective source being the corresponding edge of the grating (Fig. 8) or, more accurately, the thin luminous layer of "atmosphere" around it. In this respect, without getting into controversy outlined in [36 - 39], we can say that a certain fraction of *all combined set* of EW2 does contribute to the FF by converting into a single RW diverging from the respective edge. The fundamental feature of the FF measurement of this fraction is that, regardless of the values of *m* in (5.4), it will always give real $p_z$ and $p_x$, with $p_x^2 + p_z^2 = p^2$. The exact angular distribution of its intensity will depend on geometry of the edges. In particular, the slid waves from transmission side can progress away from the grating with a relatively large downward *z*-component of **p**, and those from the reflection side – with the large upward component, so the measured $|p_z|$ in these cases may exceed $|p_x|$. Regardless of geometry, while having arbitrarily small $\lambda_m$ interpreted formally as arbitrarily high $|p_x| > p$ in the NF, we will always observe *the whole set* as one wave with $|p_x| \leq p$ in the FF.

This allows us to evaluate the indeterminacy $\Delta p_x$ in the dispersed state (5.1). Due to the assumed even parity along the x-direction in this state, we have $\bar{p}_x = 0$, so the variance $\overline{\Delta p_x^2} = \overline{p_x^2} - \overline{p}_x^2 = \overline{p_x^2}$. Thus, in the FF measurements, the $p_x$-indeterminacy is always restricted by $\Delta p_x \leq p$, and therefore even the whole set of EW2 cannot significantly contribute to the indeterminacy associated with the RW subset in (5.1a, b). If we restrict only to that subset, then the probability for a wave with $k_x^{(m)} = mk_d$ to show up in the $k_x$-measurement is

$$\mathcal{P}_m = (2m_c + 1)^{-1} \tag{5.5}$$

Therefore

$$\overline{p_x^2} = \sum_{|m|=1}^{m_c} \left(p_x^{(m)}\right)^2 \mathcal{P}_m = 2\sum_{m=1}^{m_c} m^2 \hbar^2 k_d^2 \mathcal{P}_m = \frac{2\hbar^2 k_d^2}{2m_c + 1} \sum_{m=1}^{m_c} m^2 \tag{5.6}$$

Using the identity [40]

$$\sum_{m=1}^{m_c} m^2 = \frac{1}{6} m_c (m_c + 1)(2m_c + 1) \tag{5.7}$$

we obtain

$$\Delta p_x = \sqrt{\overline{(p_x)^2}} = \hbar k_d \sqrt{\frac{1}{3} m_c (m_c + 1)} \tag{5.8}$$

In view of (5.4), $\Delta p_x$ is a discontinuous step-like function of $p$ always remaining less than $p$.

In realistic case of a screen with finite size we must include the contribution from the EW2 that have "slid" from the screen as described above. We evaluate such contribution, assuming, in the spirit of the previous discussion, that all EW2 act as one wave with $|p_x| \leq p$ on either left or right side of the screen. In other words, since all of them are indistinguishable, we add them before squaring, assuming that their individual amplitudes and phases change on the way to the FF so that the result amounts to a single RW in (5.1b). This changes $\mathcal{P}_m = (2m_c + 1)^{-1}$ to

$$\mathcal{P}_m \to \tilde{\mathcal{P}}_m = (2m_c + 3)^{-1}, \tag{5.9}$$

and the same calculations lead to

$$\overline{p_x^2} = \left\{ \sum_{|m|=1}^{m_c} \left( p_x^{(m)} \right)^2 + 2p^2 \right\} \mathcal{P}_m = \frac{2\hbar^2 k_d^2}{2m_c + 3} \left\{ \sum_{m=1}^{m_c} m^2 + \frac{k^2}{k_d^2} \right\},$$

$$\Delta p_x = \hbar k_d \sqrt{\frac{m_c(m_c+1)(2m_c+1) + 6(k/k_d)^2}{3(2m_c+3)}} \tag{5.10}$$

If $m_c = 0$ (no crossed RW), then all $\Delta p_x$-indeterminacy is due only to EW2 and is equal to

$$\Delta p_x = \sqrt{\frac{2}{3}}\, \hbar k \tag{5.11}$$

Thus, contrary to possible naïve expectations from Fig. 2, the physical momentum indeterminacy in a dispersed state will remain finite in all FF measurements while the coordinate indeterminacy $\Delta x \sim Nd$ may be arbitrarily large at $N \to \infty$ (Fig. 9). And the measurement itself, apart from recording the input momentum $p$, will give no information about any individual $k_x$ in the EW2 states.

### 5c. Near-field measurements

Consider now the NF-measurements. In this case, the "measuring device" may be a probing beam of test particles moving along the $x$-direction as suggested in [28]. Each particle can be described as a wave packet with a Gaussian distribution over $z$:

$$\Phi(x,z) = \Phi_0 e^{-\frac{1}{2}\sigma^2(z-b)^2} e^{iKx}, \tag{5.12}$$

where $b$ is the packet's distance from the plane $z = 0$, and $K$ is the propagation number. Ideally, the particles must form a pure ensemble with the minimal possible $b$. The good candidates for absorption of quasi-tachyonic quanta of EW2 could be the fast electrons. But they will strongly interact with the surface of grating and emit Smith-Purcell radiation [41] which may mask their interaction with EW2 photons. This will greatly complicate the NF momentum measurements in EW2. The remaining candidates may be sufficiently heavy (and accordingly slow) ions or, better still, neutral atoms with appropriate velocities and optical transition frequencies.

If, despite the argument in the end of Sec. 5a, the quasi-tachyonic eigenstates (5.3) could still be singled out experimentally, the probability of absorption of $m$-th state would be proportional to

$$\tilde{\mathcal{P}}_m \sim \left| \int e^{-\frac{1}{2}\sigma^2(z-b)^2} e^{-\chi_m z} dz \right|^2 \tag{5.13}$$

For sufficiently high $m$ the corresponding eigenstates "cling" so close to the grating that the shape of the packet (5.12) beyond the distance $z \sim \chi_m^{-1}$ becomes immaterial, so we can set $\sigma = 0$ and approximate (5.13) by

$$\tilde{\mathcal{P}}_m \underset{|m|\gg m_c}{\longrightarrow} \chi_m^{-2} = \frac{1}{m^2 k_d^2 - k^2} \approx (mk_d)^{-2} \tag{5.14}$$

The *formal* $p_x$-indeterminacy turns out to be infinite

$$\Delta p_x \approx \left\{ 2\hbar^2 k_d^2 \sum_{m=m_c}^{\infty} m^2 \tilde{\mathcal{P}}_m \right\}^{1/2} \to \infty , \tag{5.15}$$

and the total indeterminacy $\Delta p_x \Delta x$ at $N \to \infty$ would spread over the whole phase space. For a grating with aperture function (4.1) ($N$ finite), the Fourier-transform of $\Psi(x)$ is a continuous function of $k_x$

$$\mathcal{F}(k_x) = \frac{1}{\sqrt{2\pi}} \frac{\sin(N k_x d / 2)}{\sin(k_x d / 2)} e^{-\frac{1}{2}i(N+1)k_x d} , \tag{5.16}$$

which is consistent with its WF being discrete along $x$ and continuous along $p_x$. To get the probability distribution in this case, we multiply (5.14) by $|\mathcal{F}(k_x)|^2$ and change $mk_d \to k_x$, $\tilde{\mathcal{P}}_m \to d\tilde{\mathcal{P}}(k_x)/dk_x$:

$$\frac{d\tilde{\mathcal{P}}(k_x)}{dk_x} \sim \frac{1}{2\pi} \frac{\sin^2(N k_x d / 2)}{k_x^2 \sin^2(k_x d / 2)} , \quad |k_x| \gg k \tag{5.17}$$

The $k_x$-spectrum, while getting continuous, retains its range unbounded, and is only modulated by periodic factor $|\mathcal{F}(k_x)|^2$. Therefore one could expect an infinite indeterminacy $\Delta p_x$ in this case as well. Indeed, if all $k_x$ were observable in the NF measurements, we would obtain for $\Delta p_x$ an expression similar to (5.15):

$$\Delta p_x \sim \left\{ 2\hbar^2 \int_k^{\infty} k_x^2 \, d\tilde{\mathcal{P}}(k_x) \right\}^{1/2} \to \infty \tag{5.18}$$

But according to the above-presented argument, even though Eq-s (5.15, 18) are formally correct, the quantity $p_x^{(m)}$ (or $p_x$) cannot be interpreted as physically observable momentum *in preserved environment* when $m > m_c$. As emphasized above, an accurate $p_x$-measurement cannot be performed close to the surface of the grating. The restrictions for possibilities of measuring individual $p_x^{(m)}$ in EW2 in the NF arise from the analysis of energy-momentum exchange between EW and environment. For instance, one cannot extract any information about $p_x$ by measuring momentum of a probing particle after it has absorbed an EW-photon: due to conservation of momentum, such particle will itself become evanescent, and the problem of measurement will just switch from one object to another. Generally, all interactions in the NF involve the particle-grating entanglement and EW-state transfer to a probing beam. These phenomena are interesting in their own right and will be considered in a separate article.

But in any case, the output state *before the measurement* may be represented by an arbitrarily large area $\Delta p_x \Delta x >> (1/2)\hbar$ in the phase space "strewn" with the arbitrarily small "cells" $\delta p_x^{(m)} \delta x_n << (1/2)\hbar$. In the idealized model (2.1), (2.4) the area spreads over the whole phase space while getting totally discrete (each cell shrinks to a point).

*Summary*

An apparently abstract concept of dispersed indeterminacy describes some familiar phenomena including multiple interference and EW2, rich in their behavior and physical properties. In the considered approach (direct application of the Wigner function formalism to a discontinuous state (2.1)), the state becomes truly discrete in both – coordinate and momentum spaces in the limit $a \to 0$, $N \to \infty$. The $p_x$-momentum measurements of a single photon in EW2 state fall into two categories: FF and NF measurements. The FF measurements will always give the transverse momentum $|p_x| \leq p$. And there is a compelling argument presented in the article against the possibility of the NF observation of $p_x > p$ for EW2. Whether this argument alone is totally sufficient still remains, to my knowledge, an open question requiring an additional analysis of the energy-momentum exchange between EW2 and probing particles.

**Figures**

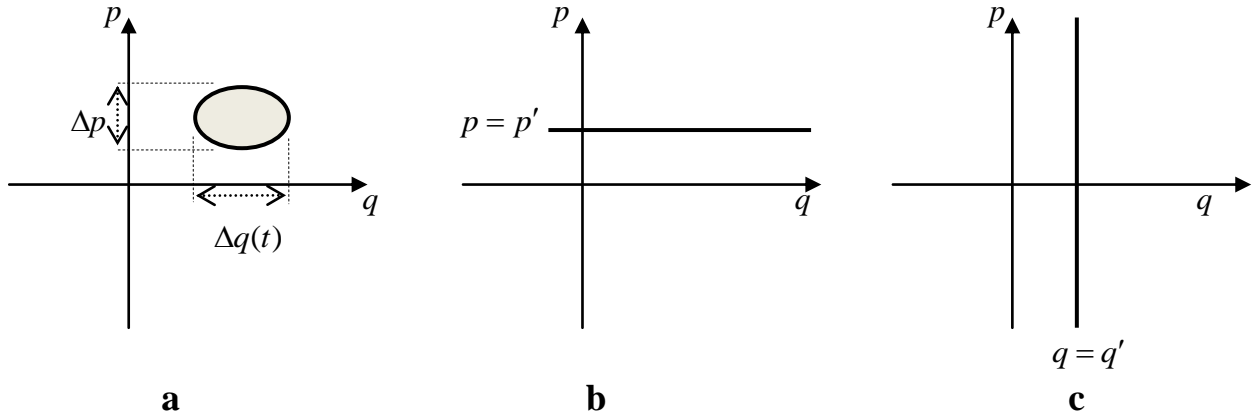

**Fig. 1**
Granular phase space

Incompatibility of observables $p$ and $q$ is reflected in the granular structure of the corresponding phase space: in contrast with Classical Mechanics (CM), a QM state can be represented by a cell with the least possible area $\Delta A_{min} = (1/2)\hbar$, rather than by a single point on the phase diagram. The shape of the cells is generally not specified and depends on a state of the system.

(a) A possible state of a particle described by a wave packet; whereas $\Delta p$ is a fixed characteristic of the given packet, $\Delta q(t)$ and thereby $\Delta A$ is generally a function of time (the packet's shape evolves); the cells with *fixed minimal area* $(1/2)\hbar$ represent the coherent states of a quantum oscillator; (b) A state with sharply defined $p = p_0$ (de Broglie's wave); (c) A state with sharply defined $q = q_0$ (instantaneously localized particle).

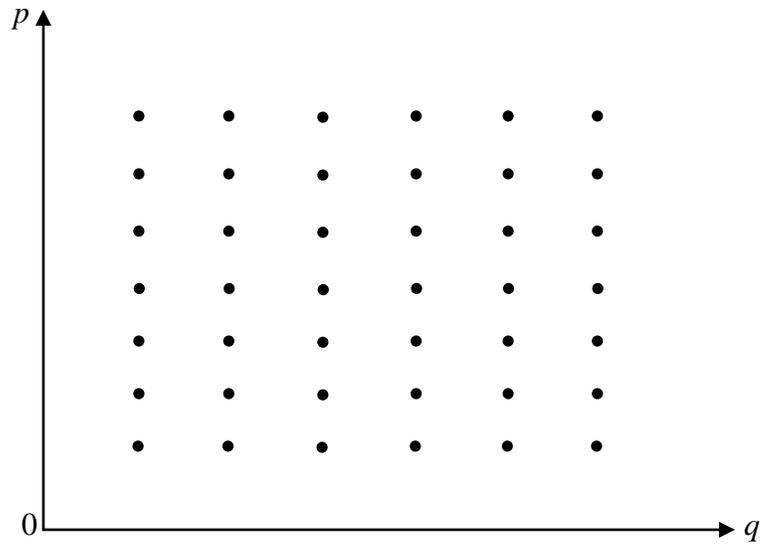

**Fig. 2**
Dispersed indeterminacy in the phase space.
The whole set of dots here belongs to one state. The net area of all dots exceeds $(1/2)\hbar$ even if the area of each dot may be vanishingly small. Even when each dot shrinks to a point, the position indeterminacy $\Delta q \to \infty$ at $N \to \infty$, and the product $\Delta p \, \Delta q \to \infty$.

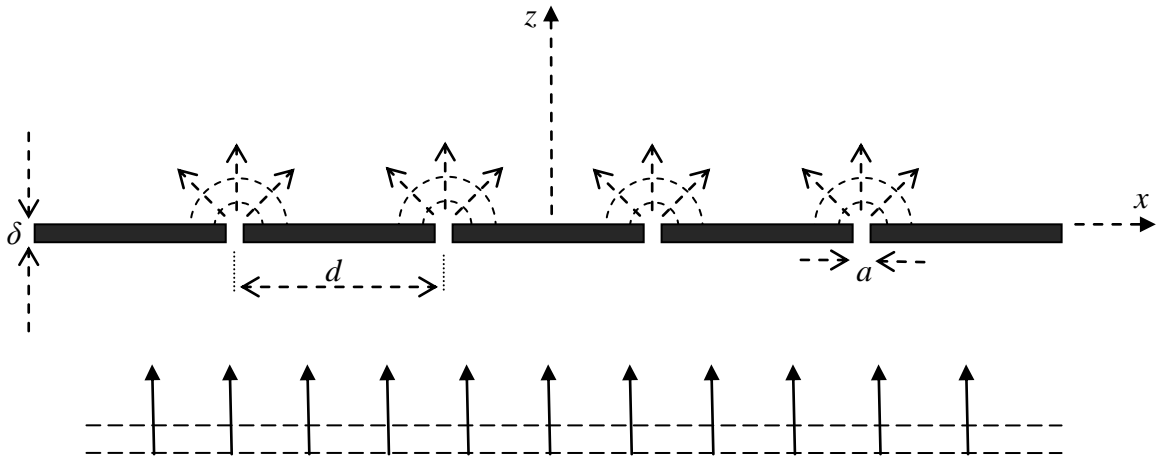

**Fig. 3**
Schematic of the multiple interference experiment with diffraction grating

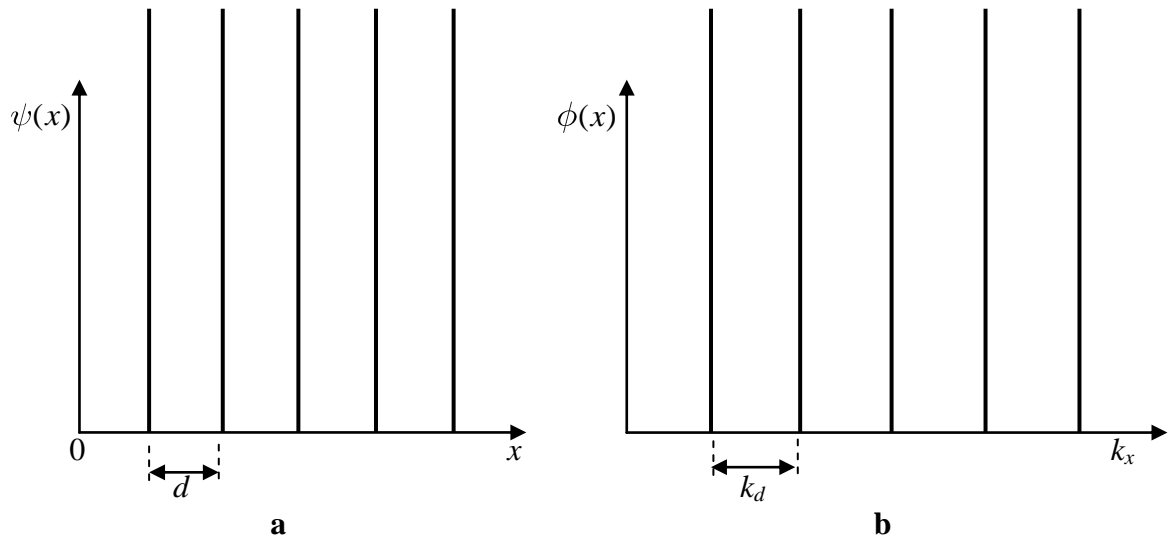

**Fig. 4**
An idealized model of the aperture function of a grating (the comb-function) in a
multiple-interference experiment
(a) - in configuration space;   (b) - in momentum space
(the number of slits is assumed to be infinite)

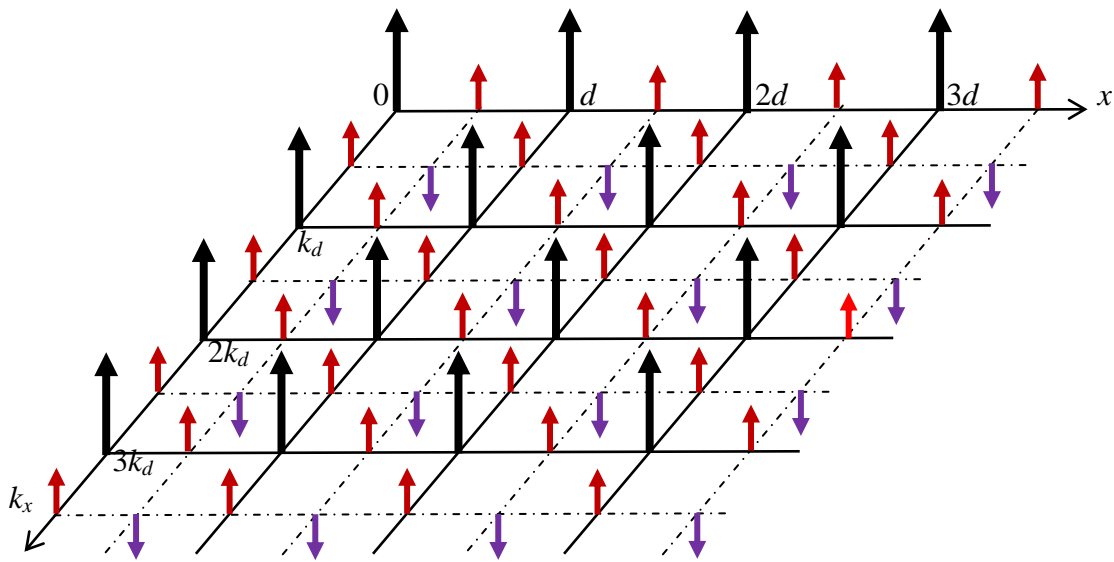

**Fig. 5**
The Wigner distribution function for a dispersed state in the limit $N \to \infty$

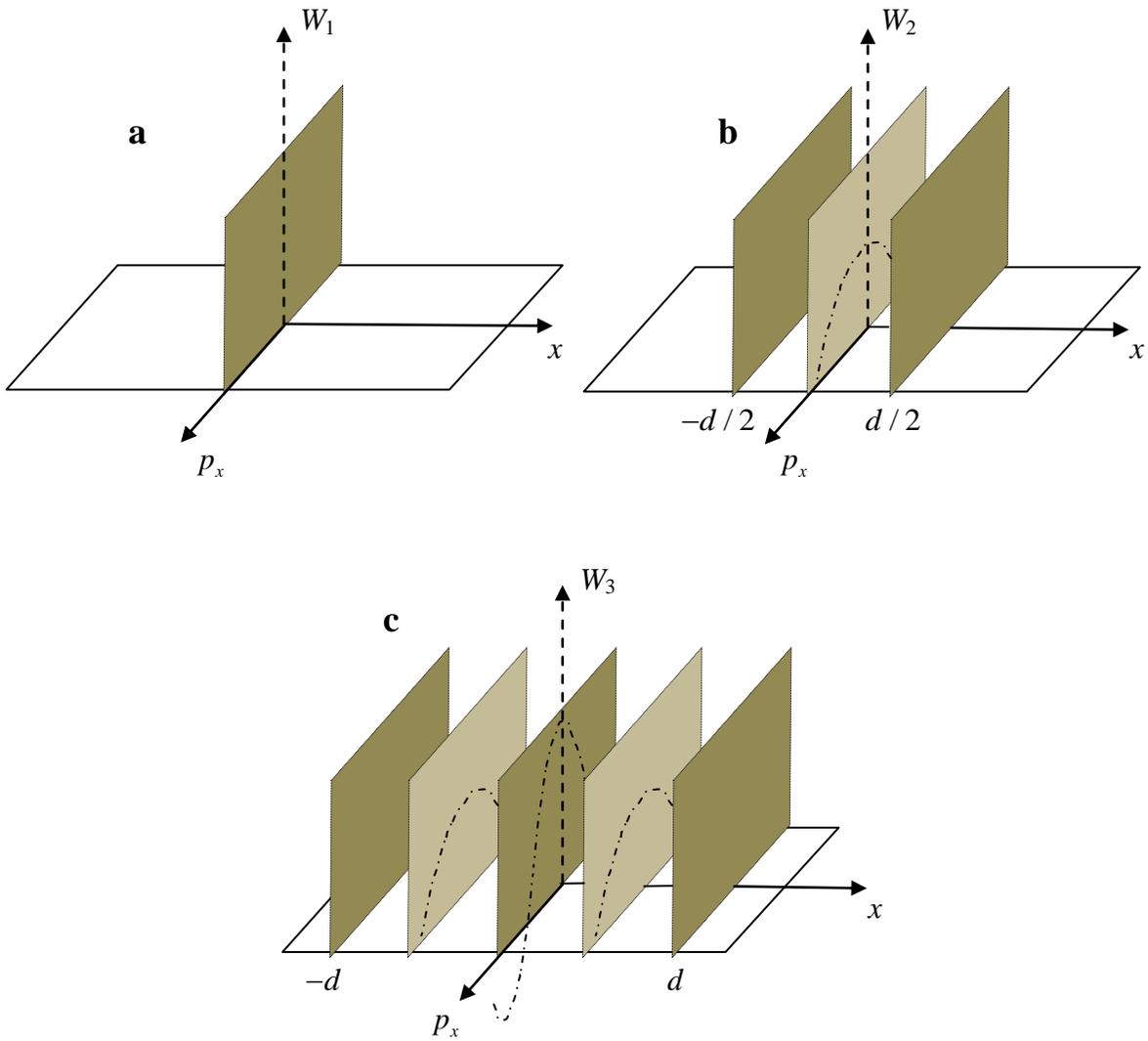

**Fig. 6**
WF for a grating with $a \to 0$ and finite $N$.
In all such cases, the WF is "semi-discrete" (discrete in the $x$-dimension and continuous in the $p_x$-dimension). The dash-dotted curves (not to scale) represent the functions "modulating" the corresponding $\delta$-functions along their baselines. Averaging over the period of a curve eliminates the "ghost" terms between the slits and leaves all other terms positive.
(**a**) $N = 1$;  (**b**) $N = 2$;  (**c**) $N = 3$

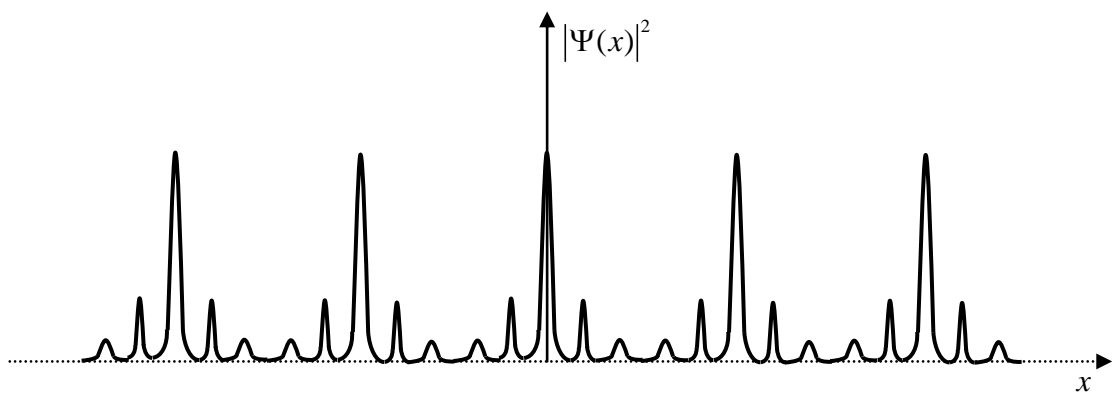

**Fig. 7**
The "post-dispersed" output state (state at some $z > 0$) as a function of $x$.

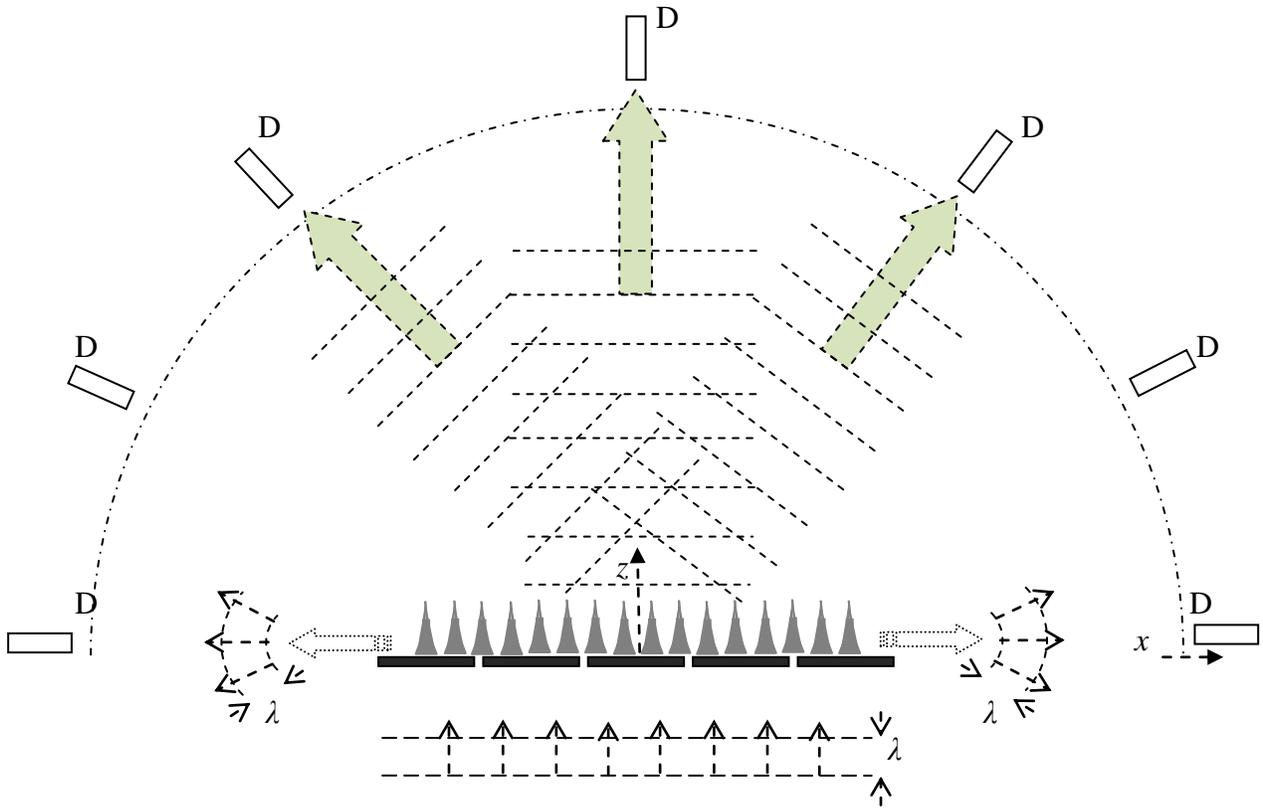

**Fig. 8**
The RW and EW in the output generated by the dispersed state within the grating. To the left and right of the grating, all set of EW converts into a single RW with the same wavelength λ as in the input wave.

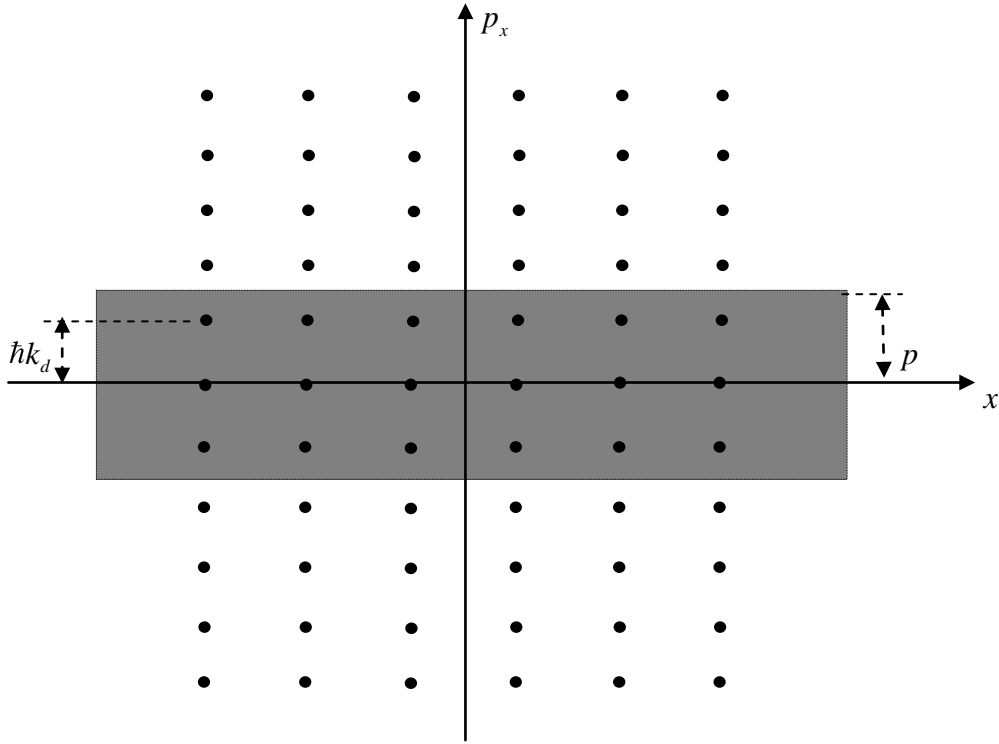

**Fig. 9**
The spread indeterminacy in the phase space.
The whole set of dots here belongs to one state. The net area of all dots exceeds $(1/2)\hbar$ even if the area of each dot may be vanishingly small. The shaded region represents indeterminacies in the coordinate and momentum dimensions, respectively. The *physical* momentum indeterminacy $\Delta p_x \sim p$ is finite in all FF and in all NF measurements. The position indeterminacy $\Delta x \to \infty$ in the limit $N \to \infty$.